\documentclass{emulateapj}
\usepackage{color}
\usepackage{ulem}

\shorttitle{ECSNe as sources of $^{60}$Fe}

\shortauthors{Wanajo et al.}

\begin{document}

\title{Electron-capture supernovae as sources of $^{60}$F\lowercase{e}}

\author{Shinya Wanajo\altaffilmark{1},
        Hans-Thomas Janka\altaffilmark{2},
        and 
        Bernhard M\"uller\altaffilmark{2}
        }

\altaffiltext{1}{National Astronomical Observatory of Japan,
        2-21-1 Osawa, Mitaka, Tokyo 181-8588, Japan;
        shinya.wanajo@nao.ac.jp}

\altaffiltext{2}{Max-Planck-Institut f\"ur Astrophysik,
        Karl-Schwarzschild-Str. 1, D-85748 Garching, Germany}

\begin{abstract}
We investigate the nucleosynthesis of the radionuclide $^{60}$Fe in
electron-capture supernovae (ECSNe). The nucleosynthetic results are based
on a self-consistent, two-dimensional simulation of an ECSN as well as
models in which the densities are systematically
increased by some factors (low-entropy models). 
$^{60}$Fe is found to be appreciably made in neutron-rich ejecta during the
nuclear quasi-equilibrium phase with greater amounts being produced in the
lower-entropy models. Our results, combining them with the yields of
core-collapse supernovae (CCSNe) in the literature, 
suggest that ECSNe account
for at least 4--30\% of live $^{60}$Fe in the Milky Way. ECSNe
co-produce neutron-rich isotopes, $^{48}$Ca, $^{50}$Ti, $^{54}$Cr, some
light trans-iron elements, and possibly weak $r$-process elements
including some radionuclides such as $^{93}$Zr, $^{99}$Tc, and
$^{107}$Pd, whose association with $^{60}$Fe might have been imprinted
in primitive meteorites or in the deep ocean crust on the Earth.
\end{abstract}


\section{Introduction}\label{sec:intro}

The origin of the radionuclide $^{60}$Fe \citep[halflife of
2.62~Myr,][]{Rugel2009} has been extensively discussed in connection to
gamma-ray astronomy \citep[an overview of the subject can be obtained
from][]{Diehl2011}. The 1173~keV and 1332~keV emission from $^{60}$Fe
decay has been confirmed by the space-based telescopes RHESSI
\citep{Smith2004} and INTEGRAL/SPI \citep{Harris2005}, indicating
ongoing nucleosynthesis of $^{60}$Fe in the Milky Way \citep[for recent
reviews, see][]{Prantzos2010, Diehl2013}. The sources of $^{60}$Fe have
generally been associated with massive stars and subsequent
core-collapse supernovae (CCSNe), in which successive neutron captures
on Fe isotopes create $^{60}$Fe \citep{Timmes1995, Huss2009}. However,
recent CCSN nucleosynthesis calculations \citep{Rauscher2002,
Limongi2006, Woosley2007} predict the ratio of $^{60}$Fe to $^{26}$Al
(halflife of 0.717~Myr) being several times greater than the line flux
ratio inferred from the INTEGRAL/SPI experiment, $^{60}$Fe/$^{26}$Al$=
0.148 \pm 0.06$ \citep{Wang2007}. \citet{Prantzos2004} suggested that
the discrepancy could be alleviated if the dominant $^{26}$Al
contributors were Wolf-Rayet star winds that did not eject $^{60}$Fe.

A detection of live $^{60}$Fe in the deep ocean crust on the
Earth has also been recently reported \citep{Knie2004, Fitoussi2008}, which may
be a sign of $^{60}$Fe injection from a nearby supernova (SN) into the
heliosphere a few Myr ago \citep{Fields2005, Fields2008}. The origin of
live $^{60}$Fe in the early solar system has been continuously
discussed since its discovery in primitive meteorites
\citep{Tachibana2003, Mostefaoui2005, Bizzarro2007}. The initial ratio
at the solar birth, $^{60}\mathrm{Fe}/^{56}\mathrm{Fe} \sim 6 \times
10^{-7}$ \citep[e.g.,][]{Mishra2010}, appeared to be higher than the
interstellar-medium (ISM) value, $\sim 3 \times 10^{-7}$
\citep{Huss2009, Tang2012}.\footnote{This value ignores the
(highly uncertain) evolution of $^{60}$Fe from the solar birth to the
present day in the Milky Way \citep[see][]{Huss2009}.} This fact led to
an idea that one or several nearby SNe had injected freshly synthesized
$^{60}$Fe into the early solar system \citep{Wasserburg1998,
Boss2013}. A recent meteorite study suggests, however, an initial ratio
of $^{60}\mathrm{Fe}/^{56}\mathrm{Fe} \sim 1 \times 10^{-8}$ \citep[see
also][]{Moynier2011, Telus2012}, which is 30 times lower than the ISM
value. If this is true, the live $^{60}$Fe might have been simply
inherited from the ISM to the molecular cloud that made the solar system
after a certain decay interval \citep[$\sim 15$~Myr,][]{Tang2012}. This
assumption, however, needs a mechanism to avoid $^{60}$Fe coming from
CCSNe during that period of time \citep[see, e.g.,][]{Gounelle2012}.
\citet{Vasileiadis2013} suggested that the low
$^{60}\mathrm{Fe}/^{56}\mathrm{Fe}$ ratios were not representative of
the proto-solar values.

It should be noted that $^{60}$Fe production in CCSN models is subject to
uncertainties in several reaction rates \citep{Woosley2007, Tur2010} as
well as in the treatment of mass loss, convection, explosion energy, and
initial metallicity in stellar models \citep{Limongi2006,
Woosley2007}. The calculated $^{60}$Fe yields should thus be taken with
caution. A possible solution to the aforementioned conflicts with
observations would thus be that CCSNe actually produced little $^{60}$Fe
and other sources with longer stellar lifetimes supplied the Galactic
$^{60}$Fe. Such sources could be asymptotic-giant-branch \citep[AGB, with
a C-O core,][]{Wasserburg2006} or super-AGB \citep[SAGB, with an O-Ne-Mg
core,][]{Lugaro2012} stars, and high-density thermonuclear SNe
\citep[SNe~Ia,][]{Woosley1997}. 

In this Letter, we report that
electron-capture SNe \citep[ECSNe,][]{Nomoto1987, Kitaura2006,
Wanajo2009}, a sub-class of CCSNe\footnote{In this Letter, the use of
``CCSNe'' is restricted to Fe-core-collapse SNe only.}  arising
from SAGB stars, can be additional sources of $^{60}$Fe in the Milky
Way. We adopt our recent nucleosynthesis results of \citet{Wanajo2013a}
and show that $^{60}$Fe is produced in appreciable amounts in the
neutron-rich and low-entropy ejecta.

\section{ECSN model and nucleosynthesis}\label{sec:model}

We employ the nucleosynthesis results of \citet{Wanajo2013a}, which are
briefly summarized below. The nucleosynthesis analysis made use of 2000
representative tracer particles, by which the thermodynamic histories of
ejecta chunks were followed in our 2D hydrodynamic calculation of an
ECSN \citep{Janka2008, Wanajo2011}. Our ECSN model predicts the
core-ejecta mass of $1.14 \times 10^{-2} M_\odot$ with electron
fractions (number of protons per nucleon) of $Y_\mathrm{e} \approx
0.40$--0.55 and entropies of $s \approx
13$--25~$k_\mathrm{B}$~nucleon$^{-1}$ \citep[$k_\mathrm{B}$ is the
Boltzmann constant; see Fig.~1
in][]{Wanajo2013a}\footnote{Throughout this Letter $Y_\mathrm{e}$ 
and $s$ are evaluated when the temperatures drop to 5~GK.}.
Post-processing nucleosynthesis calculations with an
up-to-date reaction network code \citep[with the reaction library
REACLIB version 2.0,][]{Cyburt2010} predict interesting production of
light trans-iron elements \citep[and presumably weak $r$-process
elements,][]{Wanajo2011}, whose astrophysical origin has not been fully
resolved \citep[see, e.g.,][]{Wanajo2013b}. A neutron-rich isotope,
$^{48}$Ca, whose origin remains a long-standing mystery of
nucleosynthesis \citep{Meyer1996, Woosley1997}, is also found to be made
in the neutron-rich ejecta with $Y_\mathrm{e} \approx 0.40$--0.42 and $s
\approx 13$--15~$k_\mathrm{B}$~nucleon$^{-1}$.

In addition to their ``unchanged'' ECSN model, \citet{Wanajo2013a} also
explored models in which the densities were increased by multiplying
a constant scaling factor $f$ for all the tracer particles (``$\rho
\times f$''). This effectively decreased the entropy by the same
factor. It was found that increasing the densities by factors
of 1.3 or 2 ($f = 1.3$ or 2, corresponding to
a reduction by a factor of 1.3 or 2 in entropy) leads
to a remarkable enhancement of the $^{48}$Ca abundance. This is a
consequence of the fact that a reduction of the entropy turns the
nucleosynthesis condition from $\alpha$-rich QSE (nuclear
quasi-equilibrium) to $\alpha$-poor QSE. In the latter condition, an
upward-$A$ shift of the heavy abundances in the QSE cluster is
suppressed owing to the paucity of light particles (neutrons, protons,
and $\alpha$'s). As a result, $^{48}$Ca at the low-$A$ tip of the QSE
cluster survives.
In this Letter, we
also analyze these low-entropy models.

\section{$^{60}$F\lowercase{e} production in ECSN\lowercase{e}}\label{sec:fe60}

\begin{figure}
\epsscale{1.0}
\plotone{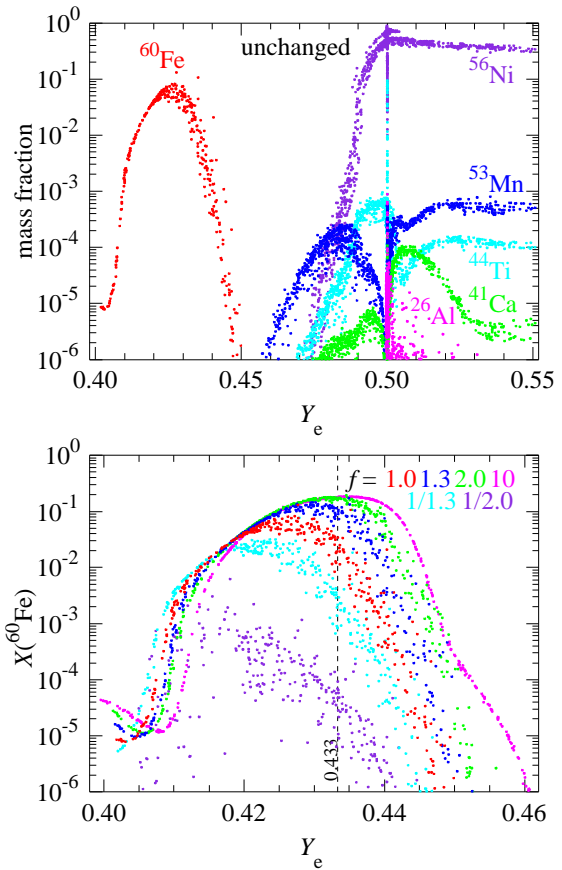}

\caption{Top: Final mass fractions of radionuclides $^{26}$Al,
$^{41}$Ca, $^{44}$Ti, $^{53}$Mn, $^{60}$Fe, and $^{56}$Ni for all the
tracer particles of the unchanged ECSN model as functions of
$Y_\mathrm{e}$.  Bottom: Final mass fractions of $^{60}$Fe for the
tracer particles in the range of $Y_\mathrm{e} < 0.462$. Also indicated
by a dashed line is $Y_\mathrm{e, nuc} = 0.433$. The result of the
unchanged model ($f = 1$) is shown in red, and those with the densities
multiplied by scaling factors $f= 1.3$, 2.0, 10, 1/1.3, and 1/2.0 are
given in different colors.}

\label{fig:yefe60}
\end{figure}

\begin{deluxetable}{ccccccc}
\tablecaption{Radioactive Yields (in units of $10^{-5} M_\odot$)}
\tablewidth{0pt}
\tablehead{
\colhead{model} &
\colhead{$^{26}$Al} &
\colhead{$^{41}$Ca} &
\colhead{$^{44}$Ti} &
\colhead{$^{53}$Mn} &
\colhead{$^{60}$Fe} &
\colhead{$^{56}$Ni}
}
\startdata
unchanged &   0.00439   &   0.0196   &   0.206  &   0.111  &  3.61 & 293 \\
$f = 1.3$ &   0.00231   &   0.0156   &   0.193  &   0.108  &  7.71 & 340 \\
$f = 2.0$ &   0.00119   &   0.00806  &   0.155  &   0.0975 &  13.0 & 405 \\
CCSNe\tablenotemark{a} &   4.69 &   2.10     &   1.52   &   26.5   & 10.4 & 10800 \\
CCSNe\tablenotemark{b} &   5.45 & --- & --- & --- &  8.31 & ---
\enddata
\tablenotetext{a}{IMF-averaged CCSN yields, adopting the solar metallicity
 models of 15--$25 M_\odot$ stars in \citet{Rauscher2002}.}
\tablenotetext{b}{IMF-averaged CCSN yields for $^{26}$Al and $^{60}$Fe,
 adopting the solar metallicity models of 12--$120 M_\odot$ stars in
 \citet{Brown2013}.}
\end{deluxetable}

The final mass fractions 
 of $^{60}$Fe are shown in Figure~\ref{fig:yefe60} (top) as functions of
$Y_\mathrm{e}$ along with those for other astrophysically important
radionuclides, $^{26}$Al, $^{41}$Ca, $^{44}$Ti, $^{53}$Mn, and
$^{56}$Ni. Among these species only $^{60}$Fe forms in the most
neutron-rich investigated conditions with $Y_\mathrm{e} \approx
0.40$--0.45, which is somewhat isolated from $Y_\mathrm{e} \approx
0.46$--0.55 in which the others are produced. These isotopes are made in
NSE (nuclear statistical equilibrium) and QSE, and, in part, by $\alpha$
and proton captures after the QSE freezeout.
The smaller core-ejecta mass of an ECSN results in
several 10 times smaller amounts of these isotopes (1st line in Table~1)
than in CCSNe \citep[4th and 5th lines in Table~1, in which the
abundances taken from][are mass-averaged by the stellar initial mass function,
IMF; see Section~\ref{sec:galaxy}]{Rauscher2002, Brown2013}.

Despite the small core-ejecta mass, we find a similar amount of
$^{60}$Fe for ECSNe comparable to that for CCSNe. This is due to appreciable
production of $^{60}$Fe in QSE with neutron-rich conditions for
$Y_\mathrm{e} \sim Y_\mathrm{e, nuc} = 26/60 = 0.433$ (characterizing
the structure of $^{60}$Fe), which is absent in CCSN ejecta. In fact,
$^{60}$Fe is the most tightly bound isotope in the range $Y_\mathrm{e,
nuc} < 0.438$.
The mass fraction $X(^{60}\mathrm{Fe})$, however, peaks at
$Y_\mathrm{e} = 0.428$, which is slightly below 0.433 (red dots in
Figure~\ref{fig:yefe60}, bottom). This is due to the presence of a more
tightly bound isotope $^{64}$Ni 
 with $Y_\mathrm{e,
nuc} = 0.438$. 

\begin{figure*}
\epsscale{1.0}
\plotone{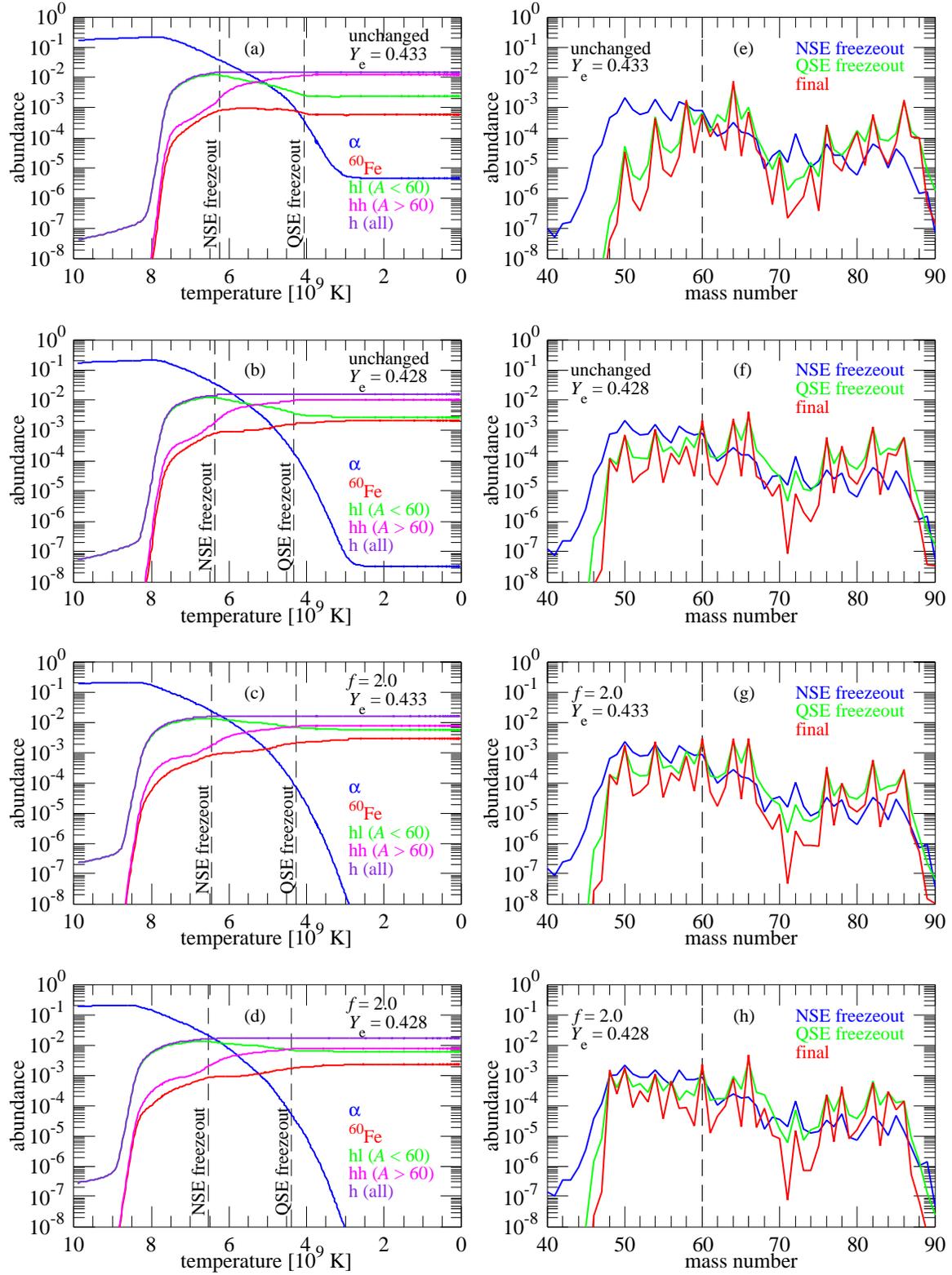}

\caption{Left: Abundances of $\alpha$, $^{60}$Fe, and heavy nuclei (for
$A < 60$, $A > 60$, and all the range) as functions of descending
temperature for the tracer particles with $Y_\mathrm{e} = 0.433$ ((a)
and (c)) and 0.428 ((b) and (d)) of the unchanged model ((a) and (b))
and those for $f = 2.0$ ((c) and (d)). The long-dashed lines mark the
NSE-freezeout and QSE-freezeout temperatures. Right: Nuclear abundances
at the NSE freezeout, at the QSE freezeout, and at the end of
calculations for the same tracer particles ((e)--(h)). The dashed line
in each panel marks the position of $^{60}$Fe.}

\label{fig:tabun}
\end{figure*}

Figure~\ref{fig:tabun} elucidates the nuclear evolutions for two
representative tracer particles with $Y_\mathrm{e} = 0.433$ (a) and
0.428 (b). The entropies are $s = 14.9\, k_\mathrm{B}\,
\mathrm{nucleon}^{-1}$ and $13.6\, k_\mathrm{B}\,
\mathrm{nucleon}^{-1}$, respectively. The expansion timescales, defined
as the $e$-folding times of the temperature drop below 0.5~MeV, are
$\tau_\mathrm{exp} = 63.8$~ms and 61.8~ms, respectively. The abundances
(number per nucleon, $Y \equiv X/A$) of $\alpha$, $^{60}$Fe, and heavy
nuclei (``h'', $A > 4$) are drawn as functions of descending
temperature. Also shown are the abundances of heavy nuclei with $A < 60$
(``hl'') and with $A > 60$ (``hh''). We find that the heavy abundance,
$Y_\mathrm{h}$, approaches a constant value around 6~GK.
This is a freezeout from NSE, defined
here when the timescale of heavy abundance formation, $\tau_\mathrm{h}
\equiv Y_\mathrm{h}/\dot{Y_\mathrm{h}}$, exceeds $\tau_\mathrm{exp}$.

We realize an upward-$A$ shift of the heavy abundances after the NSE
freezeout from decreasing $Y_\mathrm{hl}$ and increasing $Y_\mathrm{hh}$
in Figures~\ref{fig:tabun}(a) and (b). This is a result of the
$\alpha$-rich freezeout from NSE \citep{Woosley1992} followed by QSE
\citep{Meyer1998}, recognized by $Y_\alpha/Y_\mathrm{h} = 2.57$ and 2.33
at the NSE freezeout for the $Y_\mathrm{e} = 0.433$ and 0.428 cases,
respectively. We
define the QSE freezeout when the
timescale of the abundance formation for $A > 60$, $\tau_\mathrm{hh}
\equiv Y_\mathrm{hh}/\dot{Y}_\mathrm{hh}$, exceeds
$\tau_\mathrm{exp}$. QSE freezes out typically around 4~GK \citep{Meyer1998}
and the upward-$A$ shift of the heavy abundances ceases.

We find in Figures~\ref{fig:tabun}(a) and (b) that the $^{60}$Fe
abundances for $Y_\mathrm{e} = 0.433$ and $Y_\mathrm{e} = 0.428$,
respectively, decrease and increase during the QSE phase.
Figures~\ref{fig:tabun}(e) and (f) clarify the reason, illustrating the
nuclear abundances at the NSE freezeout, at the QSE freezeout, and at
the end of calculation for each tracer particle. We find that, at the
NSE freezeout, $^{60}$Fe belongs to the lighter group of the NSE
cluster.

For the $Y_\mathrm{e} = 0.433$ case (Figures~\ref{fig:tabun}(a) and
(e)), a drastic upward-$A$ shift of the heavy abundances takes place
during the QSE phase. As a result, a part of the $^{60}$Fe abundance is
taken by the heavier group, in particular by $^{64}$Ni.
For the $Y_\mathrm{e} = 0.428$ case (Figures~\ref{fig:tabun}(b) and
(f)), the upward-$A$ shift is smaller as a result of the smaller
$Y_\alpha/Y_\mathrm{h}$ at the NSE freezeout. More importantly, the
$Y_\mathrm{e}$ is appreciably smaller than the $Y_\mathrm{e, nuc}$ of
$^{64}$Ni, making $^{60}$Fe the most tightly bound isotope in this
condition. As a result, $^{60}$Fe keeps increasing in the QSE cluster
and even after the QSE freezeout.

In summary, $^{60}$Fe forms in NSE and further increases or decreases in
QSE depending on the neutron-richness as well as the available number of
$\alpha$'s during the QSE phase. The latter condition is closely related
to entropy. In the following, we thus inspect the ECSN models in which
densities are multiplied by a scaling factor $f$ for all the tracer
particles.

Figure~\ref{fig:yefe60} (bottom) shows the final mass fractions of
$^{60}$Fe for the unchanged model ($f = 1$) and those with $f = 1.3$,
2.0, 10, 1/1.3, and 1/2.0. We find a strong sensitivity of the $^{60}$Fe
production to entropy.
The nuclear evolutions for $f = 2.0$ are presented in
Figure~\ref{fig:tabun}(c) for $Y_\mathrm{e} = 0.433$ and in
Figure~\ref{fig:tabun}(d) for 0.428. The $Y_\alpha/Y_\mathrm{h}$ ratios
at the NSE freezeout are 1.46 and 1.33, respectively, being only
slightly greater than unity, as a result of reduced entropies by about a
factor of 2.
As a result, an upward-$A$ shift of the abundances is restricted by a
small number of light particles. $^{60}$Fe thus survives and increases
during the QSE phase, being maximal around $Y_\mathrm{e, nuc} = 0.433$
(Figure~\ref{fig:yefe60}, bottom).

The resulting ejecta masses of radionuclides for $f = 1.3$ and 2.0 are
presented in Table~1 (2nd and 3rd lines). $^{60}$Fe 
is appreciably produced in the low-entropy
models. A decrease of only about 30\% in entropy doubles the ejecta mass of
$^{60}$Fe, being comparable to that for CCSNe. About a factor of 2
decrease in entropy leads to about 4 times greater $^{60}$Fe amount,
being already close to that for the extreme, $f = 10$ case ($1.40 \times
10^{-4} M_\odot$; not presented in Table~1). The ejecta mass of
$M_\mathrm{ej}(^{60}\mathrm{Fe}) \sim 1 \times 10^{-4} M_\odot$ can thus
be taken to be the upper limit for ECSNe. 

\section{Contribution to Galaxy and Solar System}\label{sec:galaxy}

\begin{deluxetable}{cccccc}
\tablecaption{Most Overproduced Isotopes and ECSN Contributions}
\tablewidth{0pt} \tablehead{ \colhead{model} & \colhead{isotope} &
\colhead{$X/X_\odot$} & \colhead{$f_\mathrm{ECSN}$} &
\colhead{$f_\mathrm{60Fe}$} &
\colhead{$^{60}$Fe/$^{26}$Al\tablenotemark{a}} } \startdata unchanged &
$^{86}$Kr & 355 & 0.0854 & 0.0391 & 0.0268 \\ $f = 1.3$ & $^{74}$Se &
125 & 0.165 & 0.155 & 0.121 \\ $f = 2.0$ & $^{48}$Ca & 80.9 & 0.240 &
0.332 & 0.328 \enddata \tablenotetext{a}{Number ratios by assuming
$f_\mathrm{60Fe} = 1$ (see the text).}
\end{deluxetable}

The contribution of ECSNe to the Galactic $^{60}$Fe depends on the mass
window leading to SNe from the stellar SAGB mass range
\citep{Nomoto1987, Siess2007, Poelarends2008}. From their stellar
evolution models, \citet{Poelarends2008} obtained the initial mass range
for SAGB stars to be 7.5--$9.25 M_\odot$ in the solar metallicity
case. Assuming that all this range leads to the SN channel, the fraction
of ECSNe relative to all SN events (ECSNe + CCSNe) becomes
$f_\mathrm{ECSN} = 0.253$ by adopting the Salpeter IMF ($\propto
M_\mathrm{star}^{-2.35}$) with the upper-end of $120
M_\odot$.\footnote{The result is not very sensitive to this value. The
upper-mass of $40 M_\odot$, e.g., gives $f_\mathrm{ECSN} =
0.275$.} This can be regarded as the absolute upper limit of
$f_\mathrm{ECSN}$ in the local universe with the metallicity near the
solar value.

We further evaluate the upper limit on $f_\mathrm{ECSN}$ based on our
result. For the unchanged model, the most overproduced isotope relative
to the solar value is $^{86}$Kr (Table~2, 1st line). Given that
$^{86}$Kr in the Milky Way was exclusively made by ECSNe, we have
\citep{Wanajo2011},
\begin{eqnarray}\label{eq:fecsn}
\frac{f_\mathrm{ECSN}}{1-f_\mathrm{ECSN}}
=
\frac{X_\odot(^{86}\mathrm{Kr})/X_\odot(^{16}\mathrm{O})}
{M_\mathrm{ECSN}(^{86}\mathrm{Kr})/M_\mathrm{CCSN}(^{16}\mathrm{O})},
\end{eqnarray}
where $X_\odot(^{86}\mathrm{Kr}) = 2.39 \times 10^{-8}$ and
$X_\odot(^{16}\mathrm{O}) = 6.60 \times 10^{-3}$ are the mass fractions
of these isotopes in the solar system
\citep{Lodders2003}. $M_\mathrm{ECSN}(^{86}\mathrm{Kr}) = 6.23 \times
10^{-5} M_\odot$ is the $^{86}$Kr mass for the unchanged ECSN
model. $M_\mathrm{CCSN}(^{16}\mathrm{O}) = 1.63 M_\odot$ is the
IMF-averaged $^{16}$O mass per CCSN event, in which the yields are taken
from \citet[][their Table~1]{Brown2013}.\footnote{This is a subset of
the yields from the ``A'' series in \citet{Woosley2007}.} 
With these values, we get $f_\mathrm{ECSN} = 0.0854$ for the unchanged
model. For the low-entropy models with $f = 1.3$ and 2.0,
Equation~(\ref{eq:fecsn}) gives $f_\mathrm{ECSN} = 0.165$ and 0.240,
respectively, by replacing $^{86}$Kr with the most overproduced
isotopes, $^{74}$Se and $^{48}$Ca.

Taking the IMF-averaged $^{60}$Fe mass,
$M_\mathrm{CCSN}(^{60}\mathrm{Fe})= 8.31 \times 10^{-5} M_\odot$ with
the CCSN yields in \citet{Brown2013}, the fractions of the Galactic
$^{60}$Fe from ECSNe (relative to that from all SN events) become
$f_\mathrm{60Fe} = 0.0391$, 0.155, and 0.332 for the unchanged, $f =
1.3$, and $f = 2.0$ cases, respectively. This indicates that ECSNe
supply about 4--30\% of live $^{60}$Fe in the Milky Way. It should be
noted that the ratio from the CCSN yields,
$^{60}\mathrm{Fe}/^{26}\mathrm{Al} = 0.661$, is already more than 4
times greater than the observational flux ratio of 0.148
\citep{Wang2007}. A contribution from ECSNe would thus enlarge the
discrepancy. As noted in Section~\ref{sec:intro}, however, $^{60}$Fe
production in CCSNe is subject to uncertainties in several reaction
rates as well as in astrophysical modeling of stellar evolution.
Contributions from ECSNe could therefore be greater than the above
estimate. As an extreme case, we provide the ratios of
$^{60}$Fe/$^{26}$Al with no $^{60}$Fe (but $^{26}$Al) contribution from
CCSNe (i.e., $f_\mathrm{60Fe} = 1$) in Table~2 (last column). We find
that the low-entropy model with $f = 1.3$ gives the value that is
roughly consistent with the gamma-ray observation.

If the Galactic $^{60}$Fe were exclusively produced by ECSNe, their
longer progenitor lifetimes ($> 15$~Myr)
could give rise to different distributions between $^{26}$Al and
$^{60}$Fe. On the one hand, the $^{26}$Al distribution appears to be
clumpy as evidenced by the INTEGRAL/SPI mission \citep{Diehl2013}. Some
of this clumpiness is associated with regions hosting many young, massive
stars such as the Cygnus region. On the other hand, $^{60}$Fe may not be
associated with such young stellar regions and thus be distributed more
diffusely. Although the Cygnus region marginally appears within the
INTEGRAL sensitivity for $^{60}$Fe, no signal of its decay has been
found \citep{Martin2010}. This could be due to the age of the Cygnus
complex being much younger than the lifetimes of ECSN progenitors.

The signatures of $^{60}$Fe production in ECSNe might have been
imprinted also in primitive meteorites or in the deep ocean crust. ECSNe
produce appreciable $^{48}$Ca \citep[also $^{50}$Ti and 
$^{54}$Cr,][]{Wanajo2013a} 
that cannot be made by CCSNe. Its association
with excess $^{60}$Fe could thus be a sign of the ECSN origin. In fact,
such a correlation in meteorites was reported by \citet{Chen2011}. Note,
however, that both $^{60}$Fe and $^{48}$Ca could also originate from a
rare class of high density SNe~Ia \citep{Woosley1997}. Our ECSN model,
however, produces almost all light trans-iron nuclei up to $Z = 40$
\citep[Figure~5 in][]{Wanajo2013a} and presumably weak $r$-process
nuclei up to $Z = 50$ \citep[Figure~5 in][]{Wanajo2011}. The latter can
also be created in the subsequent neutrino-driven outflows
\citep{Wanajo2013b}. The weak $r$-process products should include a few
radionuclides with lifetimes comparable to that of $^{60}$Fe, such as
$^{93}$Zr (1.53~Myr), $^{99}$Tc (0.211~Myr), and $^{107}$Pd
(6.5~Myr). Therefore, it will be crucial to find correlations also with
these trans-iron species that are not made by SNe~Ia.

\section{Summary}

We examined the production of $^{60}$Fe in ECSNe in connection to the
nucleosynthetic results of \citet{Wanajo2013a}. The models were based on
the 2D core-collapse simulation \citep{Janka2008, Wanajo2011} of an $8.8
M_\odot$ SAGB star \citep{Nomoto1987}. In addition to the unchanged ECSN
model, we adopted the low-entropy models of \citet{Wanajo2013a}, in which
densities were multiplied by a factor $f$. We found appreciable $^{60}$Fe
production during the NSE and subsequent QSE phases in the
neutron-rich ejecta with $Y_\mathrm{e} \sim 0.43$. The amount of
$^{60}$Fe is highly dependent on entropy; lower entropy models ($f = 1.3$
and 2.0) make more $^{60}$Fe.

The unchanged ECSN model predicted $\sim$4\% contribution of ECSNe
(relative to all SN events) to the Galactic $^{60}$Fe. This fraction
could increase to $\sim$30\% (for $f = 2.0$) if the low-entropy models
were adopted. If this were the case, the Galactic flux ratio of
$^{60}\mathrm{Fe}/^{26}\mathrm{Al} = 0.148$ \citep{Wang2007} would be
explained without $^{60}$Fe contributions from CCSNe. If the Galactic
$^{60}$Fe were dominantly supplied from ECSNe (i.e., the CCSN yields
were severely overestimated), $^{60}$Fe would be more diffusely
distributed than $^{26}$Al without showing clear associations with young
stellar regions such as the Cygnus complex. This should be confirmed by
future gamma-ray line surveys.

Our ECSN models co-produce the neutron-rich isotope $^{48}$Ca
\citep{Wanajo2013a}, light trans-iron elements, and possibly weak
$r$-process elements \citep{Wanajo2011, Wanajo2013b}, accompanied by
several radionuclides with million-year lifetimes (e.g., $^{93}$Zr,
$^{99}$Tc, and $^{107}$Pd). Correlations between $^{60}$Fe and these
isotopes in primitive meteorites \citep{Chen2011} or in the deep ocean
crust on the Earth \citep{Knie2004, Fitoussi2008} will be an invaluable
evidence of $^{60}$Fe production in ECSNe.

Finally, it should be cautioned that further improvements of
hydrodynamical models (e.g., three-dimensional, high-resolution, and
general-relativistic treatment) will be needed before drawing more firm
conclusions. Studies of $^{60}$Fe production by a mini s-process during
the SAGB stage \citep{Lugaro2012} prior to ECSN explosions are also
important to evaluate the net $^{60}$Fe ejecta from such stars.

\acknowledgements

S.W. was supported by the JSPS Grants-in-Aid for Scientific Research
(23224004). At Garching, support by Deutsche Forschungsgemeinschaft
through grants SFB/TR7 and EXC-153 is acknowledged.


\begin{thebibliography}{}
\bibitem[Bizzarro et al.(2007)]{Bizzarro2007}
 Bizzarro, M., Ulfbeck, D., Trinquier, A., Thrane, K., Connelly, J. N.,
 \& Meyer, B. S. 2007, Science, 316, 1178
\bibitem[Boss \& Keiser(2013)]{Boss2013}
 Boss, A. P. \& Keiser, S. A. 2013, \apj, 770, 51
\bibitem[Brown \& Woosley(2013)]{Brown2013}
 Brown, J. M. \& Woosley, S. E. 2013, \apj, 769, 99
\bibitem[Chen et al.(2011)]{Chen2011}
 Chen, H.-W., Lee, T., Lee, D.-C., Jiun-San Shen, J., \& Chen,
 J.-C. 2011, ApJL, 743, L23
\bibitem[Cyburt et al.(2010)]{Cyburt2010}
 Cyburt, R. H., Amthor, A. M., Ferguson, R., et al. 2010, ApJS, 189, 240
\bibitem[Diehl et al.(2011)]{Diehl2011}
 Diehl, R., Hartmann, D., \& Prantzos,
 N. (ed.) 2011, Astronomy with Radioactivities
 (Lecture Notes in Physics, Vol. 812; Berlin: Springer)
\bibitem[Diehl(2013)]{Diehl2013}
 Diehl, R. 2013, Rep. Prog. Phys. 76, 026301
\bibitem[Fields et al.(2005)]{Fields2005}
 Fields, B. D., Hochmuth, K. A., \& Ellis, J. 2005, ApJ, 621, 902
\bibitem[Fields et al.(2008)]{Fields2008}
 Fields, B. D., Athanassiadou, T., \& Johnson, S. R. 2008, ApJ, 678, 549
\bibitem[Fitoussi et al.(2008)]{Fitoussi2008}
 Fitoussi, C., Raisbeck, G. M., Knie, K., et al. 2008, \prl, 101, 121101
\bibitem[Gouonelle \& Meynet(2012)]{Gounelle2012}
 Gounelle, M., \& Meynet, G. 2012, A\&A, 545, A4
\bibitem[Harris et al.(2005)]{Harris2005}
 Harris, M. J., Kn\"odlseder, J., Jean, P., et al. 2005, \aap, 433, L49
\bibitem[Huss et al.(2009)]{Huss2009}
 Huss, G. R., Meyer, B. R., Srinivasan, G., Goswami, J. N., \& Sahijpal, S.
 2009, GeCoA, 73, 4922
\bibitem[Janka et al.(2008)]{Janka2008}
 Janka, H.-Th., M\"uller, B., Kitaura, F. S., \& Buras, R. 2008, \aap,
 485, 199
\bibitem[Kitaura et al.(2006)]{Kitaura2006}
 Kitaura, F. S., Janka, H.-Th., \& Hillebrandt, W. 2006, \aap,
 450, 345
\bibitem[Knie et al.(2004)]{Knie2004}
 Knie, K., Korschinek, G., Faestermann, T., Dorfi, E. A., Rugel, G.,
 Wallner, A. 2004, \prl, 93, 1103
\bibitem[Limongi \& Chieffi(2006)]{Limongi2006}
 Limongi, M. \& Chieffi, A. 2006, \apj, 647, 483
\bibitem[Lodders(2003)]{Lodders2003}
 Lodders, K. 2003, \apj, 591, 1220
\bibitem[Lugaro et al.(2012)]{Lugaro2012}
 Lugaro, M., Doherty, C. L., Karakas, A. I., et al. 2012, M\&PS, 47, 1998
\bibitem[Martin et al.(2010)]{Martin2010}
 Martin, P., Kn\"odlseder, J., Meynet, G., \& Diehl, R. 2010, A\&A, 511, A86
\bibitem[Meyer et al.(1996)]{Meyer1996}
 Meyer, B. S., Krishnan, T. D., \& Clayton, D. D. 1996, \apj, 462, 825
\bibitem[Meyer et al.(1998)]{Meyer1998}
 Meyer, B. S., Krishnan, T. D., \& Clayton, D. D. 1998, \apj, 498, 808
\bibitem[Mishra et al.(2010)]{Mishra2010}
 Mishra, R. K., Goswami, J. N., Tachibana, S., Huss, G. R., \& Rudraswami, N. G.
 2010, ApJL, 714, L217
\bibitem[Mostefaoui et al.(2005)]{Mostefaoui2005}
 Mostefaoui, S., Lugmair, G. W., \& Hoppe, P. 2005, \apj, 625, 271
\bibitem[Moynier et al.(2011)]{Moynier2011}
 Moynier, F., Blichert-Toft, J., Wang, K., Herzog, G. F., Albarede, F.
 2011, \apj, 741, 71
\bibitem[Nomoto(1987)]{Nomoto1987}
 Nomoto, K. 1987, \apj, 322, 206
\bibitem[Poelarends et al.(2008)]{Poelarends2008}
 Poelarends, A. J. T., Herwig, F., Langer, N., \& Heger, A. 2008,
 \apj, 675, 614
\bibitem[Prantzos(2004)]{Prantzos2004}
 Prantzos, N. 2004, \aap, 420, 1033
\bibitem[Prantzos(2010)]{Prantzos2010}
 Prantzos, N., 2010, in 8th INTEGRAL Workshop ``The Restless Gamma-ray
 Universe''---Integral2010, (PoS; Trieste: SISSA), 018
\bibitem[Rauscher et al.(2002)]{Rauscher2002}
 Rauscher, T., Heger, A., Hoffman, R. D., \& Woosley, S. E. 2002, \apj, 576, 323
\bibitem[Rugel et al.(2009)]{Rugel2009}
 Rugel, G., et al. 2009, \prl, 103, 072502
\bibitem[Siess(2007)]{Siess2007}
 Siess, L. 2007, \aap, 476, 893
\bibitem[Smith et al.(2004)]{Smith2004}
 Smith, D. M. 2004, in 5th INTEGRAL Workshop on the INTEGRAL Universe,
 ed. V. Schoenfelder, G. Lichti, \& C. Winkler, ESA SP, 552, 45
\bibitem[Tachibana \& Huss(2003)]{Tachibana2003}
 Tachibana, S. \& Huss, G. R. 2003, ApJL, 588, L41
\bibitem[Tang(2012)]{Tang2012}
 Tang, H., \& Dauphas, N. 2012, E\&PSL, 359, 248
\bibitem[Telus et al.(2012)]{Telus2012}
 Telus, M., Huss, G. R., Ogliore, R. C., Nagashima, K., \& Tachibana, S. 2012,
 M\&PS, 47, 2013
\bibitem[Timmes et al.(1995)]{Timmes1995}
 Timmes, F. X., Woosley, S. E., Hartmann, D. H., Hoffman, R. D.,
 Weaver, T. A., \& Matteucci, F. 1995, \apj, 449, 204
\bibitem[Tur et al.(2010)]{Tur2010}
 Tur, C., Heger, A., \& Austin, S. M. 2010, \apj, 718, 357
\bibitem[Vasileiadis et al.(2013)]{Vasileiadis2013}
 Vasileiadis, A., Nordlund, \AA, Bizzarro, M. 2013, ApJL, 769, L8
\bibitem[Wanajo et al.(2009)]{Wanajo2009}
 Wanajo, S., Nomoto, K., Janka, H.-T., Kitaura, F. S., M\"uller, B.
 2009, \apj, 695, 208
\bibitem[Wanajo et al.(2011)]{Wanajo2011}
 Wanajo, S., Janka, H.-T., \& M\"uller, B. 2011, ApJL, 726, L15
\bibitem[Wanajo et al.(2013)]{Wanajo2013a}
 Wanajo, S., Janka, H.-T., \& M\"uller, B. 2013, ApJL, 767, L26
\bibitem[Wanajo(2013)]{Wanajo2013b}
 Wanajo, S. 2013, ApJL, 770, L22
\bibitem[Wang et al.(2007)]{Wang2007}
 Wang, W., Harris, M. J., Diehl, R., et al. 2007, \aap, 469, 1005
\bibitem[Wasserburg et al.(1998)]{Wasserburg1998}
 Wasserburg, G. J., Gallino, R., \& Busso, M. 1998, ApJL, 500, L189
\bibitem[Wasserburg et al.(2006)]{Wasserburg2006}
 Wasserburg, G. J., Busso, M., Gallino, R., \& Nollett, K. M. 2006,
 \nphysa, 777, 5
\bibitem[Woosley \& Hoffman(1992)]{Woosley1992}
 Woosley, S. E. \& Hoffman, R. D. 1992, \apj, 395, 202
\bibitem[Woosley(1997)]{Woosley1997}
 Woosley, S. E. 1997, \apj, 476, 801
\bibitem[Woosley \& Heger(2007)]{Woosley2007}
 Woosley, S. E. \& Heger, A. 2007, Phys. Rep. 442, 269
\end{thebibliography}
\end{document}